\documentclass[reprint,aps,prl,amsmath,amssymb,english]{revtex4-2}

\usepackage{graphicx}
\usepackage{dcolumn}
\usepackage{bm}
\usepackage{braket}
\usepackage{adjustbox}
\usepackage[ruled,vlined,linesnumbered]{algorithm2e}
\SetKw{KwStep}{Step:}
\usepackage[colorlinks=true, linkcolor=blue, citecolor=blue, urlcolor=blue]{hyperref}

\begin{document}


\title{Exploiting floating point emulation in tensor contractions in relativistic coupled cluster theory}
\author{Maxime Le}
\email{maxime.le@univ-lille.fr}
\affiliation{CNRS, UMR 8523–PhLAM–Physique des Lasers Atomes et Molécules, Univ. Lille, F-59000 Lille, France}

\author{Ryan Benazzouk}
\affiliation{Laboratoire de Chimie et Physique Quantique, UMR 5626 CNRS
--- Universit{\'e} Toulouse III-Paul Sabatier, 118 route de Narbonne,
F-31062 Toulouse, France}

\author{Jan Brandejs} 
\affiliation{Laboratoire de Chimie et Physique Quantique, UMR 5626 CNRS
--- Universit{\'e} Toulouse III-Paul Sabatier, 118 route de Narbonne,
F-31062 Toulouse, France}
\affiliation{Institut de Recherche en Informatique de Toulouse, UMR 5505 CNRS --- Universit{\'e} Toulouse III-Paul Sabatier, 118 route de Narbonne,
F-31062 Toulouse, France}

\author{Andr\'e Severo Pereira Gomes}
\email{andre.gomes@univ-lille.fr}
\affiliation{CNRS, UMR 8523–PhLAM–Physique des Lasers Atomes et Molécules, Univ. Lille, F-59000 Lille, France}

\date{\today}
\revised{\today}

\begin{abstract}
In this manuscript, we outline a computational strategy to leverage (vendor-)optimized implementations of the Ozaki schemes I and II for floating-point emulation as part of an implementation of the Transpose-Transpose-GEMM-Transpose (TTGT) pattern for tensor contractions, and assess its potential to accelerate two of the most expensive tensor contractions appearing in relativistic coupled cluster theory in high-end accelerators such as the NVIDIA Blackwell B200 system, in which there is a strong differential in the performance of computations for double and lower precision datatypes. Our results indicate that this emulation-based TTGT strategy can provide significant speedups compared to runs without emulation, and can be competitive with state-of-the-art, (vendor-)optimized tensor contraction libraries using native double precision as compute datatype. More importantly, our scheme provides a path towards easily allowing users to tune the desired accuracy in the tensor contraction steps, including cases where precision between single and double is sufficient for the application.
\end{abstract}

\maketitle

The interaction of matter with external perturbations such as light and electric or magnetic fields provides a way to gain an understanding of the behavior of matter at the atomistic scale, through the probing of a system's molecular properties. For an accurate description of molecular properties arising from the behavior of electrons closer to the nuclei or in systems containing heavy elements, one key ingredient is the inclusion of relativistic effects (scalar relativistic, spin-orbit coupling)~\cite{dyall_introduction_2007}. Another key ingredient is the description of electron correlation effects\cite{aszabo82}. In this task, correlated many-body methods such as coupled cluster (CC) theory~\cite{Bartlett2007,Bartlett2024} with singles and doubles excitations (CCSD), while computationally expensive, are preferable to approaches such as density functional theory (DFT), due to their ability to systematically approach the exact (full configuration interaction, FCI).

The equations to be solved in relativistic CC (R-CC) theory can be expressed in terms of several tensor contractions, that is, of products of complex-valued, high-dimensional objects representing the many-body wavefunctions and the interactions between particles. In recent years, implementations of R-CCSD that can leverage accelerated and distributed memory computer architectures have been developed, for example by~\citet{Pototschnig2021}. Typically, such implementations are carried out using (complex) double-precision floating-point arithmetic, in order to ensure accurate results.

However, it has been advanced in the literature (see~\citet{Pokhilko2018} and references therein) that double precision is perhaps too strict a requirement for obtaining correlation energies with non-relativistic (NR) CC (NR-CC) theory for ground electronic states, and algorithms can be adapted to single precision and still deliver sufficiently accurate results with the advantage of increased performance. 

Apart from the downside of extensively modifying existing codebases, in today's state-of-the-art accelerated architectures, targeting single precision does not necessarily bring about as much gain as the use of even lower precision. This is illustrated in Table~\ref{tab:cuda_flops}, where one sees that moving from double to single precision one can increase performance by less than a factor of 10, whereas further reducing precision from double to FP8, INT8 or BF16 can in principle provide much larger speedups~\cite{Matsuoka}, whether the lower precision is the final target of the computation or, as discussed below, the means of emulating a higher one.

\begin{table}
\begin{tabular}{cccccc}
\hline
  & \multicolumn{2}{c}{Blackwell} & \multicolumn{2}{c}{Blackwell Ultra} & Rubin \\ 
  & B200 SXM & GB200 & B300 SXM & GB300 &  \\ 
\hline
FP8 & 4500 & 5000 & 4500 & 5000 & 17500 \\
INT8 & 4500 & 5000 & 150 & 166 & 250 \\
BF16 & 2250 & 2500 & 2250 & 2500 & 4000 \\
FP32 & 75 & 80 & 75 & 80 & 130 \\
FP64 & 37 & 40 & 1.2 & 1.4 & 33\\
\hline
\end{tabular}
\caption{Peak performance in T(F)LOP/s for different compute types in modern NVIDIA GPUs (taken from Ref.~\cite{OS-II-FP8}).}
\label{tab:cuda_flops} 
\end{table}

An alternative to considering single or double precision can be found in algorithms based on emulation. One such approach that has gained considerable attention in recent years was introduced by Ozaki and coworkers~\cite{OS-II-INT8,OS-II-FP8}, targeting the error-free representation of matrix multiplications on high-precision datatypes with low-precision ones. Descriptions in the literature of applications of the Ozaki scheme in electronic structure so far comprise DFT~\cite{Dawson2024} and DMRG~\cite{Brower2026} calculations, and outside electronic structure it has been applied to complex-valued quantum circuit simulation~\cite{Ootomo2024}. Single-precision emulation via BF16 tensor cores has recently been applied to \mbox{NR-DF-CCSD} \cite{Bayraktar}, with \mbox{$1.7–1.8\times$} speedups per iteration at FP32 accuracy, and we note a recent description of applications (based on \texttt{cuBLAS 13.4.2}) of the Ozaki schemes to Green's functions theory (GW) and NR-CCSD theory in the NVIDIA technical blog~\cite{cublas-ozaki-1-2-blog}. 

While in the context of DFT the application of the Ozaki scheme is straightforward since one only deals with matrices, for many-body approaches such as DMRG and (N)R-CC tensors (and tensor contractions) are ubiquitous and in existing applications the Ozaki schemes are applied by mapping a given tensor contraction onto a matrix product. In contrast to DMRG, in (N)R-CC there is more variability in the contractions, in terms of sizes and the nature of the tensors involved. 

To the best of our knowledge double-precision emulation has not yet been explored for R-CC. This manuscript therefore aims to take a step in addressing this gap, and explore how the Ozaki schemes can be used to speed up contractions appearing in the theory. For that, we have leveraged the Tensor Algebra Processing Primitives (TAPP) application programming interface (API) to implement tensor contractions using the Transpose-Transpose-GEMM-Transpose (TTGT) tensor contraction pattern, in which we can optionally leverage (vendor-)optimized implementations of the Ozaki schemes I and II. Our final objective is to integrate our open-source Ozaki TTGT code into the ExaCorr module~\cite{Pototschnig2021} of the DIRAC relativistic electronic structure code~\cite{Saue2020}, as it already leverages TAPP~\cite{DIRAC261} to access different computational backend for tensor contractions.

\section{Theory}

\subsection{Relativistic coupled cluster theory}

The relativistic CCSD ansatz is based on an exponential parametrization of the ground-state wavefunction, 
\begin{equation}
\ket{\text{CC}} = e^T\ket{R}
\end{equation}
involving a closed-shell reference determinant $\ket{R}$ constructed from one-particle solutions of the mean-field (Hartree-Fock) equations employing a Hamiltonian $H$ based on the 4/2-Component Dirac equation, representing one-body ($h_i$) and two-body ($g_{ij}$) interactions~\cite{dyall_introduction_2007,sikkema_molecular_2009,Saue2020} 
\begin{align}
H &= \sum^N_i h_i + \sum_{ij} g_{ij} \nonumber \\
  &= \sum_{pq} h^p_q a^\dagger_p a_q + \frac{1}{4}\sum_{pqrs} g^{pq}_{rs} a^\dagger_p a^\dagger_q a_s a_r,
\end{align}
and excitation operators $T = T_1 + T_2$, where
\begin{align}
        T_1 &= \sum_{ai}t_i^a \hat{a}^\dagger_a\hat{a}_i\\
        T_2 &= \frac{1}{4}\sum_{abij}t_{ij}^{ab} \hat{a}^\dagger_a\hat{a}^\dagger_b\hat{a}_j\hat{a}_i.
\end{align}
represent single ($T_1$) or double ($T_2$) excitation operations, which acting upon the reference determinant $\ket{R}$ will produce singly ($\ket{\mu_1} = \ket{{}^{a}_{i}}$) or doubly ($\ket{\mu_2} = \ket{{}^{ab}_{ij}} $) excited determinants (with $\bra{\mu_i} \equiv \ket{\mu_i}^\dagger$, e.g.\ $\bra{{}^i_a} = \bra{R}\hat{a}_i\hat{a}^\dagger_a $ etc).
In the expressions above and those that follow, the set of occupied indices by the letters $i, j, k, \dots$ and the set of virtual indices by $a, b, c, \dots$, and $p, q, r, s, \dots$ represent general (occupied or virtual) indices, and $\hat{a}_p, \hat{a}^\dagger_q$ represent creation and annihilation operators in second quantization. 

To obtain the amplitudes $t$, one typically solves the projected equations involving the similarity-transformed Hamiltonian $\bar{H} = e^{-T}\hat H e^{T}$:
\begin{align}
    \bra{\mu_i}\bar{H} \ket{R}&= 0,
\end{align} 
whereas the CC correlation energy is given by
\begin{align}
    E_{CC} = \bra{R}\bar{H}\ket{R}\label{cc-energy-general}.
\end{align}

Employing the Einstein summation notation, where repeated indices are contracted, the programmable expressions to solve the equations necessary to determine the CCSD singles ($t_i^a$) and doubles ($t_{ij}^{ab}$) amplitude tensors are given by equations~\ref{eq:CC_equations_singles} and~\ref{eq:CC_equations_doubles} respectively,
\begin{align}
S^{a}_{i} & =F_{i}^{a} 
+ \left(\left[H^{k}_{c}-2 \, F^{k}_{c} \right] t^{c}_{i} \right)t^{a}_{k}
+ H^{a}_{c}t^{c}_{i} 
+ H^{k}_{i}t^{a}_{k} 
\nonumber \\
&
+ H^{k}_{c}t^{ac}_{ik}
+ \frac{1}{2}  V^{ak}_{cd}\tau^{cd}_{ik}
+ \frac{1}{2}   V^{kl}_{ic}\tau^{ca}_{kl}
- V^{ak}_{ci}t^{c}_{k}
\label{eq:CC_equations_singles}
\\
S^{ab}_{ij}  
& = \frac{1}{2} V^{ab}_{cd}\tau^{cd}_{ij} 
+ P_{ij}^{-}P_{ab}^{-} \left(H^{ak}_{ci}t^{cb}_{jk}\right)
+ P_{ji}^{-} G^{k}_{j}t^{ab}_{ik} 
+ V^{ij\dagger}_{ab}
\nonumber \\
& + P^{-}_{ij}\left( V^{cj\dagger}_{ab}t^{c}_{i} \right)
+ P_{ab}^{-} G^{a}_{c}t^{cb}_{ij}
+ P^{-}_{ab}\left(C^{ka}_{ij}t^{b}_{k}\right)
+ \frac{1}{2} A^{kl}_{ij}\tau^{ab}_{kl},
\label{eq:CC_equations_doubles}
\end{align}
with $E_{CC}$ from Equation~\ref{cc-energy-general} given as
\begin{equation}
E_{CC} = F_a^i t^a_i  + \frac{1}{4} V^{ij}_{ab} \tau^{ab}_{ij}.\label{cc-energy-einsum}
\end{equation}
Here $\tau^{ab}_{ij} = t^{ab}_{ij} - t^{a}_{i}t^{b}_{j}$, $V^{pq}_{rs}$ denote antisymmetrized two-electron integrals, $P_{pq}^{-}f_{pq} = f_{pq} - f_{qp}$ a permutation operator, $F^{p}_q$ elements the Fock matrix, with all other terms corresponding to intermediate tensors that may depend on the $F$, $t$, and $V$ tensors and whose definitions are given by~\citet{Pototschnig2021}. We note that in the simpler Coupled Cluster Doubles (CCD) ansatz, which captures a sizable fraction of electron correlation, the singles amplitudes are disregarded and one only retains a simplified expression for equation~\ref{eq:CC_equations_doubles}, but the overall computational scaling remains the same for CCD and CCSD ($n_\text{occ}^2 n_\text{virt}^4$\cite{Bartlett2007}, with $n_\text{occ}, n_\text{virt}$ the number of occupied and virtual molecular orbitals respectively).

The focus on the current work will be on the first two terms in Equation~\ref{eq:CC_equations_doubles}, which appear in both CCSD and CCD methods. The first term (also referred to as the particle-particle ladder term, hereafter referred to as PP) involves the contraction of two virtual indices of a doubles amplitude tensor with those of the corresponding (four virtual indices) two-electron integral since it represents the most expensive contraction of the method in terms of operation count and involves the largest tensor in terms of memory storage \cite{Bayraktar}. The 
second term involves contractions over one occupied and one virtual indices (also referred to as the particle-hole ladder term, hereafter referred to as PH), and is the second most expensive class of contraction present.

Additionally, since the sizes of the occupied and virtual spaces are typically quite different, by investigating both the PP and PH type contractions we can have a more nuanced picture of the performance of tensor contractions than by focusing on a single type.

To have an idea of performances in actual calculations, we calculated the CCSD energy of PbS with OS-I, OS-II and cuTENSOR.

Finally, it is important to note that, unless point group symmetry can be exploited (see~\cite{Saue2020} and references therein), for relativistic Hamiltonians including spin-orbit interactions these tensors are now in general complex-valued. This is in contrast to the case of scalar relativistic and non-relativistic Hamiltonians, for which both the integrals and amplitudes are real-valued. This has as a consequence that the number of floating-point operations in the tensor contractions above is larger for relativistic Hamiltonians compared to non-relativistic ones.

\subsection{The Ozaki schemes for floating point emulation and its use in tensor contractions}

The size of the calculations introduces roundoff errors which may be greater than experimental accuracy. Using a bigger number representation is not always a solution, since it increases the computation time and memory consumption. For matrix multiplications, the Ozaki scheme (OS)\cite{Ozaki2011,OS-II-INT8,OS-II-FP8} is an elegant method to have greater accuracy while exploiting smaller, less accurate number representations. For a matrix product $C=AB$, the OS splits the mantissa of every element of the input matrices into multiple matrix splits, then computes the General Matrix Multiply (GEMM) over each split-pair and accumulates the results. 

The purpose of the Ozaki schemes is to tackle round-off/accumulation errors due to the dynamic range of matrix elements in GEMM. The OS GEMM is error-free in the sense that it allows achieving greater accuracy than the original matrix storage data type. The original Ozaki scheme was designed to emulate floating-point precision with a smaller floating-point format. Recent developments include the OS version 1 (OS-I)\cite{Ootomo2024} available in \texttt{cuBLAS} version 13~\cite{cuBLAS}. Ozaki and co-workers implemented a second version of OS (OS-II)\cite{OS-II-INT8}, tackling the quadratic scaling of OS-I and made it available as the RIKEN-developed open-source library \texttt{GEMMul8}~\cite{OS-II-repo}. More recently, OS-II has also been made available in~\texttt{cuBLAS 13.4.2} and later~\cite{cublas-ozaki-1-2-blog}.


\textbf{The Ozaki-I scheme.}  In the Ozaki-I scheme, the idea is to split the mantissa of the elements of a matrix into INT8 blocks, while saving the exponents of the larger element, in absolute value, of every row (resp. column) in A (resp. B). That is followed by performing integer GEMMs on the splits and accumulating the results. Only the final accumulation is done in double precision. The number of GEMM operations scales as $N(N+1)/2$ where $N$ is the number of splits. While it looks like OS-I has greater complexity than the original GEMM, the integer GEMM is much faster than double-precision GEMM. An appealing property of the OS-I scheme is that the relative error of the calculation is directly linked to the number of mantissa bits that are kept. For $N$ splits, the error in the GEMM result will be of the order of 2$^{-8N}$.

\textbf{The Ozaki-II scheme.} The OS-II mantissa splitting philosophy is completely different from OS-I as it exploits modular arithmetic. The input matrices are scaled to convert the original floating-point precision GEMM into a sum of INT8~\cite{Ootomo2024} or FP8~\cite{OS-II-FP8} GEMM. Then, the scaled matrices are decomposed as a sum of module remainders by the Chinese Remainder Theorem (CRT). The result matrix is obtained by accumulating the GEMM applied to the modulo remainder from the previous step. This emulation scheme needs to perform $N$ integer GEMM operations as opposed to the quadratic scaling of its predecessor. This reduction in complexity comes with a not-so-tractable error bound, even if experimentally the relative error decreases linearly with respect to the number of splits on a log scale. The authors of \texttt{GEMMul8}~\cite{OS-II-repo} have nevertheless provided recommendations to obtain a given number of precision bits, and these are reproduced for completeness in Table \ref{tab:OS-II}. It is possible to either use INT8 or FP8 in OS-II. Even if the FP8 version is more precise, the INT8 is faster on Nvidia cards whose architecture predates Blackwell. We expect the FP8 version to be more efficient on architectures such as Blackwell Ultra or Rubin.

\begin{table}
\begin{tabular}{ccccc}
\hline
\texttt{N} &  INT8-Real & INT8-Complex & FP8-Real & FP8-Complex \\
\hline
2            &     7 &         7 &    13 &     12 \\
3            &    11 &        11 &    18 &     16 \\
4            &    15 &        15 &    23 &     22 \\
5            &    19 &       19  &   28  &     27 \\
6            &    23 &       23  &   34  &     32 \\
7            &    27 &       27  &   39  &     35 \\
8            &    31 &       30  &   44  &     40 \\
9            &    35 &       34  &   48  &     46 \\
10           &   39  &       38  &   53  &     51 \\
11           &   43  &       41  &   58  &     56 \\
12           &   47  &       45  &   63  &     61 \\
13           &   51  &       48  &   68  &     66 \\
14           &   55  &       52  &   73  &     70 \\
15           &   58  &       55  &   78  &     75 \\
16           &   62  &       58  &   84  &     80 \\
17           &   66  &       62  &   89  &     85 \\
18           &   70  &       65  &   94  &    90 \\
19           &   73  &       68  &   99  &    95 \\
20           &   77  &       71  &  102  &   100 \\
\hline
\end{tabular}
\caption{Table relating associating the number of moduli \texttt{N} and the approximate effective bits
for the different OS-II backends  (taken from from the \texttt{README.md} file in the \texttt{GEMMul8} repository~\cite{OS-II-repo}.}
\label{tab:OS-II} 
\end{table}

%

\textbf{Combining OS and tensor contraction with TTGT.} Emulation formulated at the level of the tensor contraction itself, rather than of the underlying GEMM, is only beginning to be explored~\cite{Ren2026}; we have therefore leveraged existing (vendor-)optimized implementations of the OS schemes, and combined them with the Transpose-Transpose-GEMM-Transpose (TTGT) pattern for tensor contraction. We nevertheless note that there exist other algorithms for tensor contraction that are able to hide the cost of the transposition, such as GETT~\cite{Springer2018}. 

TTGT aims to take input tensors in a given tensor contraction, reorder their indices to unveil a GEMM operation, and transform the output tensor into the form of the original operation. An example of the TTGT algorithm workflow is illustrated in Algorithm~\ref{algo:ttgt_example} for the contraction $C_{abc} = A_{aij}B_{ibjc}$. 

\begin{algorithm}[H]
\label{algo:ttgt_example}
Compute $A^T_{aij} = A_{aij}$ and $B^T_{ijbc} = B_{ibjc}$\;
GEMM $C^T_{abc} = A^T_{aij}B^T_{ijbc}$\;
$C_{acb} = C^T_{abc}$\;
\caption{Example of TTGT workflow for contraction $C_{acb} = A_{aij}B_{ibjc}$.}
\end{algorithm}

The first line of the algorithm modifies the data layout of the input tensors to create transposed tensors that group bound/free indices on the same side. The bound indices should have the same order in A and B, as we will see shortly. In our example, only B needs to be transposed. Line 2 of the algorithm can be seen as the following GEMM $C^T_{IJ} = A^T_{IK}B^T_{KJ}$ by defining global indices $I=a$, $J=bc$ and $K=ij$. To get the final result $C$, another reordering of the indices in $C^T$ may be needed. 

There is not a unique TTGT scheme, given that one can arbitrarily choose valid indices for $A^T$ and $B^T$. For instance, $A^T_{aji}$ and $B^T_{jibc}$ are also valid transpose candidates. In this work, we have not investigated whether or not we employed the best transpose candidates, because we found that in practice the bulk of the execution time in the calculations with our TTGT implementation was associated with \texttt{cuBLAS} GEMM kernels. Our implementation assumes a col-major data layout and selects the transpose candidates $A^T$ and $B^T$ that keep the contiguity of the first index.

\section{Computational details}

All calculations were carried out at the DALIA system at IDRIS, an NVIDIA GB200 NLV72 system made up of B200 (Blackwell) GPUs, with 186 GB of memory available per GPU. We employed the \texttt{cuBLAS} and \texttt{cuTENSOR} versions available as part of the NVIDIA CUDA toolkit version 13 available on DALIA (namely \texttt{cuBLAS} \texttt{13.2.0.9} and \texttt{cuTENSOR} \texttt{2.4.1}). For the Ozaki scheme II (OS-II), we used a development version (commit hash \texttt{30b5b5d}, based on version \texttt{3.5.4}) of the open-source library \texttt{\texttt{GEMMul8}} developed by RIKEN ~\cite{OS-II-repo}. Unless otherwise noted, we use the INT8 backend for the OS-II scheme.

The implementation of the PH and PP contractions has been carried out using the TAPP API~\cite{tapp}. The code is publicly available at our fork~\cite{code-hackathon-2025} of the official TAPP repository~\cite{tapp-repo}.

In the case of calculations employing the TTGT backend, our implementation employed the \texttt{cuTT} library~\cite{cutt-paper,cutt-library} for the transposition steps and the \texttt{cuBLAS} library~\cite{cuBLAS} for the GEMM steps, whereas reference calculations are carried out with the \texttt{cuTENSOR} library~\cite{cuTENSOR-ref}. In our implementation of PP and PH, the calls to \texttt{cuBLAS} and \texttt{cuTENSOR} are not done directly but through their bindings to TAPP.

Our main focus will be in the comparison of the performance of tensor contractions involving tensors of double precision complex storage type (referred to in the following as C64), and in the TTGT case we will consider both the use of emulation (via the OS-I and OS-II schemes) as well as without it, in order to gauge the overhead of transposition and to provide a more direct comparison to \texttt{cuTENSOR}. For completeness, we also compare results for tensor contractions involving tensors of single-precision complex storage type (referred to in the following as C32) obtained with \texttt{cuTENSOR}. The native C32 and C64 compute datatypes were used in \texttt{cuTENSOR}. 

In addition to the use of \texttt{cuTENSOR} without emulation, we have also explored the emulation available in 
cuTENSOR \texttt{2.4.1} (released Dec. 2025), through passing the compute descriptor \texttt{CUTENSOR\_COMPUTE\_DESC\_8XINT8} to the \texttt{cutensorCreateContraction} function. We note, however, that since the \texttt{cuTENSOR 2.4.1} documentation~\cite{cuTENSOR-ref} does not describe which algorithms are implemented, in what follows we will consider it as third emulation option and refer to it as \texttt{cuTENSOR 2.4.1 emulated}. All reported cuTENSOR emulation results apply specifically to version \texttt{2.4.1}; newer versions were not evaluated as they were not supported by the hardware driver version on DALIA at the time of this study.

Unless specified otherwise, the number of occupied orbitals n$_{occ}$ will be 100 and the number of virtual orbitals n$_{vir}$ will have the values 100, 120, 140, 160, 180 and 200. These values, chosen having in mind the maximum available memory in the GPUs, are also representative of problem sizes encountered in production calculations accounting for valence correlation in prototypical heavy element systems (see for instance those investigated in~\cite{Pototschnig2021,Yuan2023,Yuan2024,Yuan2025}, in particular if the virtual space can be compressed with techniques such as frozen virtual natural orbitals~\cite{Yuan2022}. We would like to mention runtime issues when running PP benchmarks with n$_{nir} =240$ and $300$, where the GEMM \texttt{cuBLAS} emulated routines took much more time than \texttt{cuTENSOR} C64. With the help of profiling tools, we observed a switching of GEMM kernels by \texttt{cuBLAS}, from cuBLASLt (for smallerproblems sizes) to cuTLASS (for the poorly performing cases), which each utilize different parameters.

The timings presented in the discussion have been obtained from the average of wall times over hundred executions of each tensor contraction, in order to minimize artifacts due to initialization and other runtime issues. 

The real and imaginary parts ($x$) of the tensor elements in all cases have been generated with a uniform $rand(0,1)$ and Gaussian distribution $g\sim \mathcal{N}(0,1)$:
\begin{equation}
x = (rand(0,1)-0.5)e^{\phi g}. 
\end{equation}
The positive parameter $\phi$ serves to benchmark the accuracy of the GEMM operation. The higher $\phi$ is, the greater the roundoff error. In the timing benchmark, this parameter is set to 1 in the timing benchmark and to 3 in the precision benchmark. In the precision benchmark, the results are compared to cuTENSOR-emulated results.

In the GB200 architecture and for C64, the OS-I implementation in \texttt{cuBLAS} can be tuned via three main options: the mantissa bit count, the emulation strategy and the mantissa control. The first option lets the user decide the desired accuracy of the GEMM kernel and its execution time. In the discussion below, we will discuss the effect of this parameter by the number of precision digits a given mantissa bit count ensures (e.g.\ 8 precision digits for single precision, and 16 precision digits for double precision), and will use \texttt{digits} as a shorthand throughout the text. The second option signals to the library whether to run emulation whenever possible (referred to as \texttt{eager}, controlled by the \texttt{CUBLAS\_EMULATION\_STRATEGY\_EAGER} option) or only when it has performance benefits (referred to as \texttt{performant}). The third option sets the behavior of the GEMM kernels to adapt to the dynamic range of input tensors at runtime (referred to as \texttt{dynamic}) or not (referred to as \texttt{fix}), controlled by the \texttt{CUDA\_EMULATION\_MANTISSA\_CONTROL\_DYNAMIC} and \texttt{CUDA\_EMULATION\_MANTISSA\_CONTROL\_FIXED} options, respectively.

We note that while it is possible to use emulation of C32 in cuBLAS, that does not employ the OS-I scheme and therefore we do not investigate this option in this manuscript. For further details on the features of emulation currently implemented in cuBLAS, the reader is referred to the \texttt{cuBLAS} documentation. 

For the calculation of CCSD energies for the PbS molecule with and without emulation, the molecular electron repulsion integral and CCSD amplitudes tensors which serve as input tensors for the calculation of CCSD energies were obtained from calculations with a development version of the DIRAC code~\cite{Saue2020}, based on the DIRAC26 release~\cite{DIRAC261}. These calculations were carried out in C64 native precision. We employed the eXact 2-Component (X2C) Hamiltonian~\cite{ilias_infinite-order_2007}, with only the 10 electrons in the valence shell being correlated, while the virtual space was truncated to retain only orbitals whose energy was below 100 a.u. The basis set employed was the doubly-augmented triple-$\zeta$ Dyall basis sets~\cite{dyall2006relativistic,dyall2022diffuse} (denoted \texttt{d-aug-dyall.v3z} in DIRAC). We have employed the PbS experimental geometry found in the NIST webbook~\cite{nist-chemistry-webbook}, with the Pb atom located at the origin and the S atom placed at a distance of 2.286863 \r{A} along the z axis.


\section{Results}

\begin{figure}[!tbp]
  \centering
  \begin{minipage}[b]{0.5\textwidth}
    \includegraphics[width=\textwidth]{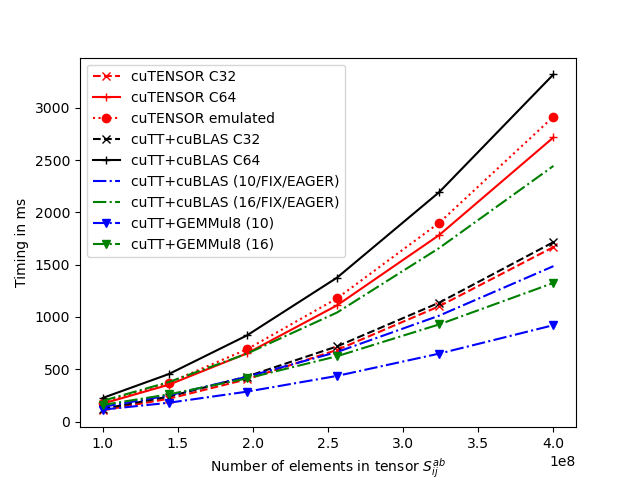}
  \end{minipage}
  \begin{minipage}[b]{0.5\textwidth}
    \includegraphics[width=\textwidth]{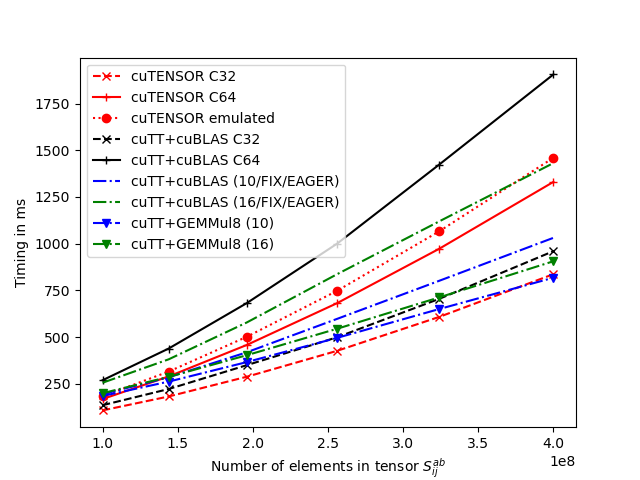}
  \end{minipage}
  \caption{Average timings (in ms) for the PP (top) and PH (bottom) contraction with respect to the number of elements in the result tensor $S_{ij}^{ab}$. The text in parenthesis gives the configuration of the calculations. For \texttt{cuTT}+\texttt{cuBLAS} backend, it is the number of precision digits, CUDA emulation control and \texttt{cuBLAS} emulation strategy. For \texttt{cuTT}+\texttt{GEMMul8}, the number in parenthesis is the number of precision digits. cuTENSOR and cuBLAS results were obtained with versions \texttt{2.4.1} and \texttt{13.2.0.9}, respectively.\label{fig:timings}}
\end{figure}

\begin{figure}[!tbp]
  \centering
  \begin{minipage}[b]{0.5\textwidth}
    \includegraphics[width=\textwidth]{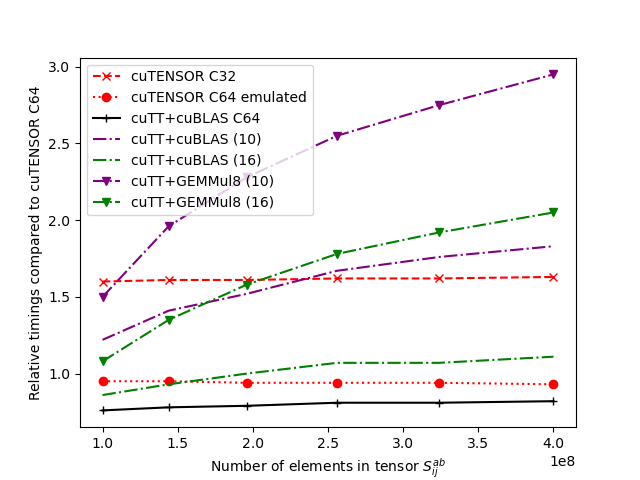}
  \end{minipage}
  \begin{minipage}[b]{0.5\textwidth}
    \includegraphics[width=\textwidth]{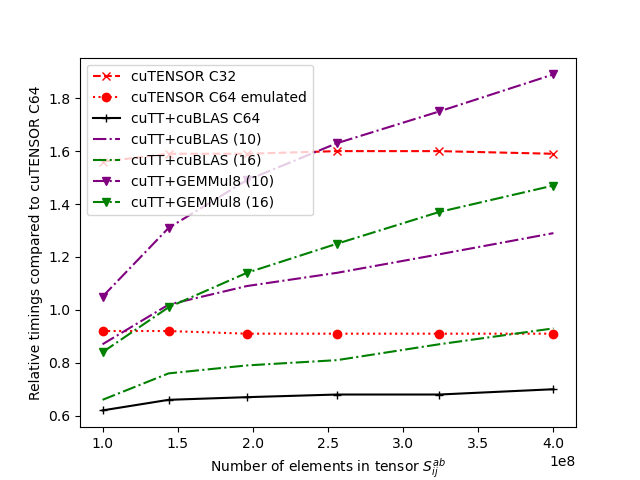}
  \end{minipage}
    \caption{Relative timings with respect to the \texttt{cuTENSOR} C64 baseline using the native C64 compute datatype (values above 1 represent speedups) for PP (top) and PH (bottom) contraction, as a function of the number of elements in the result tensor $S_{ij}^{ab}$. For OS-I we employed the \texttt{fix} and \texttt{eager} emulation parameters. cuTENSOR and cuBLAS results were obtained with versions \texttt{2.4.1} and \texttt{13.2.0.9}, respectively. \label{fig:speedups}}
\end{figure}

Our timings for PP and PH contractions are presented in Figure~\ref{fig:timings}, whereas relative performances to that of \texttt{cuTENSOR} C64 runs without emulation are presented in Figure~\ref{fig:speedups}. While in testing our Ozaki TTGT implementation with the \texttt{cuBLAS} backend we have explored different combinations of parameters, in the discussion below we will mostly focus on the combination of \texttt{eager} and \texttt{fix} options, so that we can better control for the effect of the number of \texttt{digits} on the performance.

\textbf{Performance with respect to \texttt{cuTENSOR} C64.} Focusing first on the timings in Figure~\ref{fig:timings} without emulation, we observe for both PP and PH contractions the expected trend of higher performance for single precision than for double precision for both \texttt{cuTENSOR} and TTGT. For both contractions we see a more noticeable differentiation in absolute timings between \texttt{cuTENSOR} and TTGT for C64 than for C32, which indicates that there appears to be a larger overhead due to transposition for C64 than for C32.

For the \texttt{cuTENSOR 2.4.1 emulated} results for both PP and PH contractions, we have not observed performance gains over those without emulation, but rather a small and systematic reduction in performance for all contraction sizes. This may be due to the overhead resulting from the definition of the number of retained mantissa bits at runtime in order to ensure C64 or better accuracy, as indicated by the \texttt{cuTENSOR 2.4.1} documentation.

\begin{figure}[!tbp]
  \centering
  \begin{minipage}[b]{0.5\textwidth}
    \includegraphics[width=\textwidth]{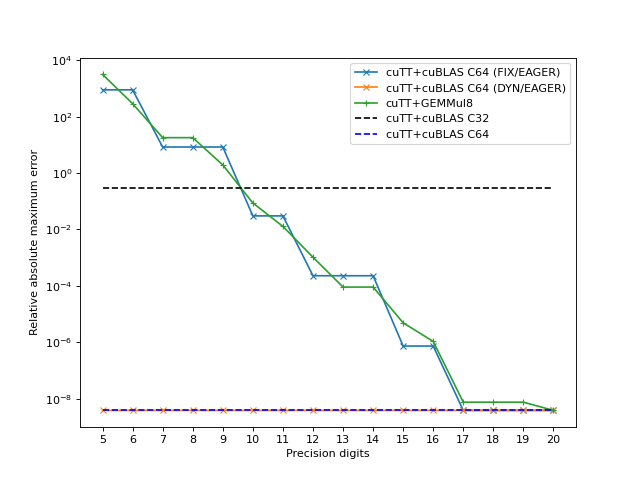}
    \caption{Maximum absolute relative error with respect to the number of \texttt{digits} in the PH contraction (using n$_\text{occ}=10$ and n$_\text{vir}=100$) for the \texttt{fix} and \texttt{dynamic} \texttt{cuBLAS} emulation configurations in TTGT. The \texttt{eager} strategy was enabled in all calculations in order to enforce emulation.}
  \end{minipage}
  \label{fig:precision}
\end{figure}

For OS-I based TTGT (\texttt{cuBLAS} backend), on the other hand, once we enable emulation and retain 16 \texttt{digits}, which is comparable to native C64 precision (though, as shown in Figure~\ref{fig:precision}, with 16 \texttt{digits} the precision may in practice be slightly lower than native C64), we have that for the PP contraction TTGT C64 starts to approach \texttt{cuTENSOR} for as the size of contractions increases, and slightly outperforms it for the largest contraction size considered. For the PH contraction \texttt{cuTENSOR} outperforms TTGT for most contraction sizes, though the performance of TTGT approaches that of \texttt{cuTENSOR} for the two largest largest contraction sizes.

When retaining 10 \texttt{digits} (which, from Figure~\ref{fig:precision}, suggests a comparable but slightly better accuracy than native C32 precision) we observe that OS-I based TTGT (\texttt{cuBLAS} backend) slightly underperforms \texttt{cuTENSOR} for smaller contraction sizes for both PP and PH contractions, but starts to outperform it for the larger sizes, while for the PH contraction the crossover occurs only at the largest size considered. 

For the OS-II based TTGT (\texttt{GEMMul8} backend), we observe consistently better performance than the OS-I based TTGT (\texttt{cuBLAS} backend). For instance, the performance for 16 \texttt{digits} with the \texttt{GEMMul8} backend is typically comparable to that of \texttt{cuTENSOR} C32, and for 10 \texttt{digits}OS-II outperforms both \texttt{cuTENSOR} C32 and OS-I TTGT for 10 \texttt{digits}. We note, however, that both \texttt{cuTENSOR} and \texttt{cuBLAS} C32 references use native single precision compute. With BF16-based emulation of single precision~\cite{Bayraktar}, \texttt{cuBLAS} C32 curves would be expected to move further up, so they should not be read as an upper bound for single precision performance.

Here it is interesting to see in more detail how the change in the number of \texttt{digits} changes the performance for the different contraction types and sizes. For that we focus on the relative performance with respect to \texttt{cuTENSOR} C64 runs, presented in Figure~\ref{fig:speedups}. From that we see that TTGT without emulation is slower than the reference (speedup factors ranging from 0.76 to 0.82 for PP, and from 0.62 to 0.70 for PH, with the best performances taking place for larger problem sizes), whereas \texttt{cuTENSOR 2.4.1 emulated} is only a few percent slower and shows very consistent performance across the different problem sizes, with speedup factors ranging from 0.95 to 0.93 for PP, and from 0.92 to 0.91 for PH. We note that \texttt{cuTENSOR} C32 speedup factors range from 1.60 to 1.63 for PP and 1.56 to 1.59 for PH, and are also quite consistent across problem sizes. In contrast, for the OS schemes there is an increase in performance with increased system size. 

For OS-I with 16 \texttt{digits} speedup factors range from 0.86 to 1.11 for PP (the smallest two problem sizes show factors of 0.86 and 0.93 respectively), and from 0.66 to 0.93 for PH, while for 10 \texttt{digits} speedup factors go up to about 1.22 to 1.83 for PP, and 0.87 to 1.29 for PH (and only the smallest problem shows a slowdown for PH). For OS-II performances are slightly better, with speedup factors ranging from 1.08 to 2.05 for PP, and from 0.84 to 1.47 for PH (only the smallest problem shows a slowdown for PH) for 16 \texttt{digits}, and with 10 \texttt{digits} the speedup factors increase to 1.50 to 2.95 for PP, and 1.05 to 1.89 for PH.

The trends discussed above are also found for other numbers of \texttt{digits} between 16 and 6 (see supplementary information for details), and as expected the speedups increase as the number of \texttt{digits} decreases.

For example, for OS-I using 12 \texttt{digits} we obtained speedup factors ranging from 1.02 to 1.42 for PP, and from 0.76 to 1.08 for PH, whereas for 6 \texttt{digits} speedup factors reach 1.66 to 3.22 for PP, and 1.10 to 1.77 for PH. Likewise, for OS-II with 12 \texttt{digits} we obtained speedup factors ranging from 1.02 to 1.42 for PP, and from 0.76 to 1.08 for PH, whereas for 6 \texttt{digits} speedup factors range from 1.96 to 4.04 for PP, and from 1.23 to 2.29 for PH.

In view of both our OS-I and OS-II results we can conclude, first, that even when retaining a precision (nearly) equivalent to C64, emulation provides an effective alternative to native C64 tensor contractions. Second, emulation provides a straightforward path to achieving increased performance with accuracies which are intermediate between C32 and C64. 

This is particularly interesting in the case of codes for which storage and contractions are natively C64 (such as DIRAC), as it suggests that it may not be necessary to carry out extensive modifications to the code in order to work with C32 storage as well, with adaptations being circumscribed to tensor operations only.

\textbf{Performance with respect to machine precision.} Even if the precision of the OS-I has been well studied\cite{Ootomo2024}, there is no clear relation between \texttt{cuBLAS} emulation options and the actual runtime behavior. We illustrate this point by comparing the performance for different numbers of \texttt{digits} together with the \texttt{fix} and \texttt{dynamic} configurations, while keeping the \texttt{eager} mode active.

In Figure~\ref{fig:precision} we show the relative maximum absolute error with respect to the number of \texttt{digits}. For the OS-I TTGT (\texttt{cuBLAS} backend) the \texttt{fix} configuration exhibits a predictable decreasing error behavior, allowing one to achieve intermediate precision between C32 and C64 by choosing an appropriate number of \texttt{digits}. A similar behavior was observed for OS-II TTGT (\texttt{GEMMul8} backend), meaning that in this case the recommendations from the RIKEN group, summarised in Table~\ref{tab:OS-II}, are sound. One can notice C32 results can be recovered at around 9-10 precision digits.


Given the improvement in performance for the PP and PH contractions discussed above as one reduces the number of \texttt{digits} from 16 to 8, if the application does not require the full precision afforded by C64 it can be advantageous to choose an intermediate value, and our scheme provides a straightforward way to do so. The ability to choose the number of \texttt{digits} provides a key diagnostic tool: whereas the precision needed by a correlated calculation has so far been assessed by comparing separate double and single precision codes \cite{Pokhilko2018}, here the same number can be imposed on all contractions, with storage kept in C64, and lowered step by step from a double precision-equivalent setting until the amplitudes, the correlation energy or the property of interest start to degrade, followed by a term-by-term analysis on which contractions set the precision requirements.
 
 It is also interesting to note that upon increasing the number of \texttt{digits} beyond 16, the emulation in principle allows one to gain additional precision beyond C64; this may offer an interesting pathway for realizing simulations in which precision higher than C64 is required, such as when taking quantum electrodynamics (QED) corrections into account~\cite{Benazzouk2026}. For that, the OS-II scheme seems to provide slightly better precision with 20 \texttt{digits}.

When changing from \texttt{fix} to the \texttt{dynamic} configuration, according to the \texttt{cuBLAS} documentation\cite{cuBLAS} it is not necessary for the user to set parameters to achieve double precision, and the algorithm adapts to the dynamic range of the inputs. This automatic tuning comes at the cost of extra processing. 

\begin{figure}[!tbp]
  \centering
  \begin{minipage}[b]{0.5\textwidth}
    \includegraphics[width=\textwidth]{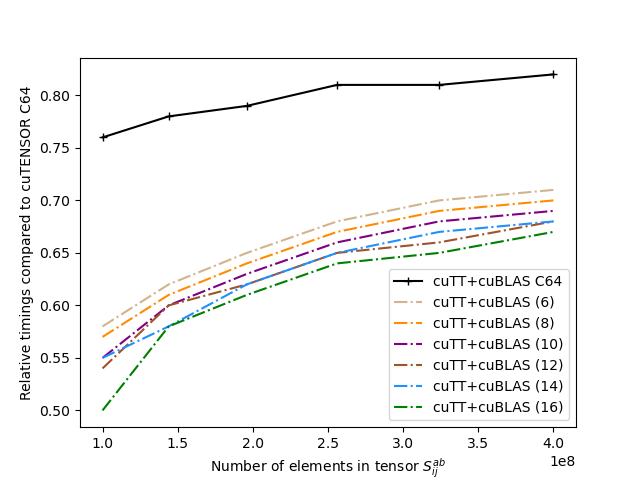}
  \end{minipage}
    \caption{Relative timings with respect to the \texttt{cuTENSOR} C64 baseline using the native C64 compute datatype (values above 1 represent speedups) for the PP contraction with respect to the number of elements in the result tensor $S_{ij}^{ab}$ for emulated TTGT using the \texttt{dynamic} setup.\label{fig:dynamic-overhead}}
\end{figure}

In our case, as can be seen in Figure~\ref{fig:dynamic-overhead} for the PP contraction, the runtime performance was negatively impacted and resulted in consistently worse performance than both the \texttt{cuTENSOR} reference (over 30\% slower, corresponding to scaling factors ranging from 0.45 to 0.57 for 16 \texttt{digits} and from 0.49 to 0.60 for 6 \texttt{digits}) and TTGT without emulation (we note \texttt{cuTENSOR 2.4.1 emulated} shows a very similar behavior as TTGT with \texttt{dynamic} configuration). Due to that, such configuration does not appear to be advantageous and should likely be avoided in the context of tensor contractions.

\subsection{Impact of precision on the R-CC correlation energy}
 
As a first step in assessing the impact of emulation, and the use of intermediate precision in R-CC calculations, in this section we evaluate how the number of \texttt{digits} affects the calculation of the correlation energy (Equation~\ref{cc-energy-einsum}) for different problem sizes, using input tensors (molecular integrals and CC amplitudes) calculated in double precision. While this comparison does not provide us with information on the impact of the error propagation in the iterative solution of the R-CCSD amplitude equations, it nevertheless serves to illustrate the impact of precision in a key step in production calculations.

Before proceeding, one should keep in mind that energy differences of about 1 Kcal/mol, typically considered the accuracy necessary to achieve chemical accuracy~\cite{Jensen2007}, correspond to energy differences of 10$^{-3}$ atomic units (a.u.). With that in mind, we argue the error introduced by emulation can be deemed acceptable in practical applications targeting chemical accuracy, if the absolute difference in energies with respect to native C64 precision is of about 10$^{-4}$ to 10$^{-4}$ a.u.\ or better.

\begin{figure}[!tbp]
  \centering
  \begin{minipage}[b]{0.5\textwidth}
    \includegraphics[width=\textwidth]{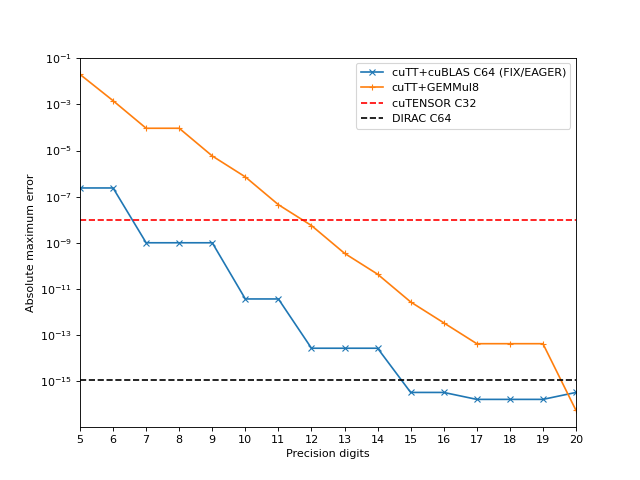}
  \end{minipage}
   \caption{Absolute errors in the closed-shell R-CCSD correlation energy for the PbS molecule (in atomic units, a.u., with n$_\text{occ}=10$ and n$_\text{vir}=400$) for and \texttt{cuTENSOR} C32 and different number of \texttt{digits} for the OS-I (\texttt{cuBLAS} backend) and OS-II (\texttt{GEMMul8} backend) based TTGT calculations, with respect to the baseline using the native C64 compute datatype in DIRAC.  Absolute values under 10$^{-4}$ to 10$^{-5}$ a.u.\ represent accuracies consistent with achieving chemical accuracy.\label{fig:energy-accuracy-comparison}}
\end{figure}

Our results for the R-CCSD energy PbS molecule are presented in Figure~\ref{fig:energy-accuracy-comparison}. In our calculations, we account for scalar relativistic and and spin-orbit effects, and for this system both effects are essential for a qualitative description of the electronic structure. Our correlation space, comprising the electrons in the valence shell (n$_\text{occ}=10$), and correlated a sufficient number of virtual orbitals (n$_\text{vir}=400$), has been used in a study of CCSD dipole polarizabilities and first hyperpolarizabilities~\cite{XS-LE}, and we have verified that it ensures that both ground-state energies and the molecular properties under consideration were converged with respect to the virtual space truncation.

We observe that with both OS schemes the absolute error (compared to DIRAC calculation employing C64 native precision) in the R-CCSD energy decreases with increased precision, though in this example we find that the OS-I based TTGT (\texttt{cuBLAS} backend, using the \texttt{fix} and \texttt{eager} parameters) seems to arrive at the desired precision somewhat faster than the OS-II based TTGT (\texttt{GEMMul8} backend), though in the latter we see a monotonic increase in the accuracy while for the former precision increases in a more stepwise manner. We see that, with both schemes, the absolute error reaches values we consider consistent with chemical accuracy for a number of \texttt{digits} roughly equivalent to single precision (8), and for OS-I a comparable accuracy to that of \texttt{cuTENSOR} C32 can be achieved with even fewer \texttt{digits}. In view of these results, we consider that emulation with the OS schemes provides users with fine control of the desired accuracy for the problem at hand.

\section{Conclusion}

In this work, we present an implementation and a benchmark study of floating-point emulation for tensor contractions central to molecular many-body electronic structure methods such as relativistic coupled cluster theory, whose tensors are typically stored in double precision for reasons of accuracy. To the best of our knowledge, this work represents the first application of floating-point emulation to relativistic coupled cluster theory, and more generally the first use of both Ozaki-I and Ozaki-II type schemes for double and intermediate precision emulation in complex-valued coupled cluster tensor contractions.

Our implementation, based on the TTGT tensor contraction pattern, extends the TAPP API in order to leverage the efficient tensor transposition implementation based on the \texttt{cuTT} library, the Ozaki-I scheme implementation in the NVIDIA \texttt{cuBLAS} (version \texttt{13.2.0.9}) library, and the Ozaki-II scheme implementation in the \texttt{GEMMul8} library (version \texttt{3.5.4}) for the GEMM step. 

From our benchmark studies using NVIDIA B200 GPUs we observe that, for double precision types without the use of floating-point emulation, calculations with the TTGT pattern are somewhat slower than the reference calculations using the \texttt{cuTENSOR} library (version \texttt{2.4.1}, employing the native C64 compute datatype), for the two classes of contractions investigated (speedup factors of about 0.8 for PP and 0.7 for PH). For \texttt{cuTENSOR} version \texttt{2.4.1}, emulation in \texttt{cuTENSOR} does not result in speedups compared to the native implementation for both contractions (speedup factors of about 0.9 for PP and PH). 

With emulation in \texttt{cuBLAS} and using the \texttt{fix} and \texttt{eager} configurations, on the other hand, we observe that our TTGT code can perform similarly to the reference for the largest problem sizes (speedup factors of about 1.1 for PP and 0.9 for PH) when targeting an accuracy corresponding to double precision (16 \texttt{digits}).  For 16 \texttt{digits}, with the emulation in \texttt{GEMMul8} we obtained somewhat better performance for the largest problem sizes (speedup factors of about 2.1 for PP and 1.5 for PH), and also observe comparable performances to the reference for the smallest problem sizes considered (speedup factors of about 1.1 for PP and 0.8 for PH).

Our scheme provides a straightforward way to target intermediate precision (e.g.\ between single and double precision, or between double and higher precision). For target accuracies between single and double precision (10, 12 or 14 \texttt{digits}), we observe an overall improvement in performance, so that TTGT can approach (and sometimes exceed) the performance of \texttt{cuTENSOR} C32 (speedup factors of about 1.6 for PP and PH compared to C64). For instance, for 12 \texttt{digits} and the largest problem size, OS-I with \texttt{fix} and \texttt{eager} configurations shows speedup factors of about 1.4 for PP and 1.1 for PH, whereas OS-II shows speedup factors of about 2.5 for PP and 1.7 for PH. 

While further reducing the target accuracy in emulation brings about more significant speedups in tensor contractions (e.g. for 8 \texttt{digits} OS-I with \texttt{fix} and \texttt{eager} configurations can arrive at speedup factors of about 2.4 for PP and 1.5 for PH for the largest problem size, whereas the respective OS-II speedups are about 3.6 for PP and 2.1 for PH), this gain in performance must be weighed against the final accuracy required for the molecular properties of interest.  In the case of the \texttt{cuBLAS} OS-I implementation, since the \texttt{fix} configuration applies a uniform, adjustable precision to every contraction, it also provides a practical means of determining the required precision empirically. 

As a first example of adjusting the accuracy to fit the application's needs, we have investigated the effect of the choice of precision on the calculation of the R-CCSD correlation energy for the PbS molecule, starting from input tensor in double precision. We demonstrated that both OS schemes enable us to keep errors within bounds comparable to the need to achieve chemical accuracy. Such a procedure can bee straightforwardly applied to computationally more expensive operations, such as the calculation on non-iterative higher-order corrections to the correlation energy. Another natural extension of this procedure will be to vary the mantissa bit count along the CC iterations, starting at reduced precision and raising it as the amplitudes converge, in the spirit of the adaptive-precision SCF of \citet{Huang2026}. 

We have observed that using the \texttt{dynamic} instead of the \texttt{fix} configuration in emulation resulted in lower performances than TTGT without emulation, and an analogous runtime behavior may be the cause for the comparable performance of \texttt{cuTENSOR} \texttt{2.4.1} with and without emulations. This suggests a need to carefully evaluate the impact of emulation strategies prior to use in production runs.


Finally, by virtue of being an extension to TAPP, our implementation can be adopted by codes already using the API, and straightforwardly extended to other architectures and Ozaki scheme implementations. The latter may become necessary if INT8 throughput is further reduced on upcoming GPU generations (Table~\ref{tab:cuda_flops})~\cite{OS-II-FP8,Matsuoka}. Integration into the DIRAC program package as a demonstrator in production calculations is currently under way.

\section{Acknowledgements}

JB and RB acknowledge support from the ERC project HAMP-vQED (grant agreement No 101019907 [01/10/2021-30/09/2026]). The work of JB was supported by the France 2030 NumPEx Exa-SofT (ANR-22-EXNU0003) and Exa-DI (ANR-22-EXNU0006) projects managed by the French National Research Agency (ANR). ML and ASPG acknowledge funding from projects CPER WaveTech, Labex CaPPA (Grant No. ANR-11-LABX-0005-01), ANR SCREECHES (Grant ANR-24-CE29-0904), and the University of Lille cross-disciplinary initiatives C2EMPI and AREA. We acknowledge the use of resources at IDRIS (Grant  AD010817866) and technical support from IDRIS (Simplice Donfack) and NVIDIA (Ryan Reese, Nicolas Blin) researchers during the May 2026 IDRIS Open Hackathon, support from the French national supercomputing facilities (Grants A0190801859, AD010818002, A0200817520) and the HPC center at the University of Lille. This research used resources of the Oak Ridge Leadership Computing Facility, which is a DOE Office of Science User Facility supported under Contract DE-AC05-00OR22725 (allocation CHP109). We also acknowledge helpful discussions with Trond Saue and Lucas Visscher in connection to the use of emulation in the DIRAC code, and with Anthony Scemama, Alfredo Buttari and \"{O}rs Legeza concerning the Ozaki schemes.

\bibliography{ms}

\end{document}



\title{Supplementary Information: Exploiting floating point emulation in tensor contractions in relativistic coupled cluster theory}
\author{Maxime Le}
\email{maxime.le@univ-lille.fr}
\affiliation{CNRS, UMR 8523–PhLAM–Physique des Lasers Atomes et Molécules, Univ. Lille, F-59000 Lille, France}

\author{Ryan Benazzouk}
\affiliation{Laboratoire de Chimie et Physique Quantique, UMR 5626 CNRS
--- Universit{\'e} Toulouse III-Paul Sabatier, 118 route de Narbonne,
F-31062 Toulouse, France}

\author{Jan Brandejs} 
\affiliation{Laboratoire de Chimie et Physique Quantique, UMR 5626 CNRS
--- Universit{\'e} Toulouse III-Paul Sabatier, 118 route de Narbonne,
F-31062 Toulouse, France}
\affiliation{Institut de Recherche en Informatique de Toulouse, UMR 5505 CNRS --- Universit{\'e} Toulouse III-Paul Sabatier, 118 route de Narbonne,
F-31062 Toulouse, France}

\author{Andr\'e Severo Pereira Gomes}
\email{andre.gomes@univ-lille.fr}
\affiliation{CNRS, UMR 8523–PhLAM–Physique des Lasers Atomes et Molécules, Univ. Lille, F-59000 Lille, France}

\date{\today}
\revised{\today}

\maketitle

\section{Definitions}

In the manuscript we refer to maximum relative errors (MREs) and maximum absolute error (MAEs) in the discussion of numerical precision and on the accuract of R-CCSD energies calculated with emulation.

The MRE  has been computed with this formula:
\begin{align}
    \text{MRE} &= \max_i \{\frac{|C^{\text{reference}}_i-C^\text{result}_i|}{|C^{\text{reference}}_i|}\}.
\end{align}
whereas the MAE has been computed with this formula:
\begin{align}
    \text{MAE} &= \max_i \{|C^{\text{reference}}_i-C^\text{result}_i|\}.
\end{align}

\section{Performance with respect to cuTENSOR}

\begin{figure}[!tbp]
  \centering
  \begin{minipage}[b]{0.75\textwidth}
    \includegraphics[width=\textwidth]{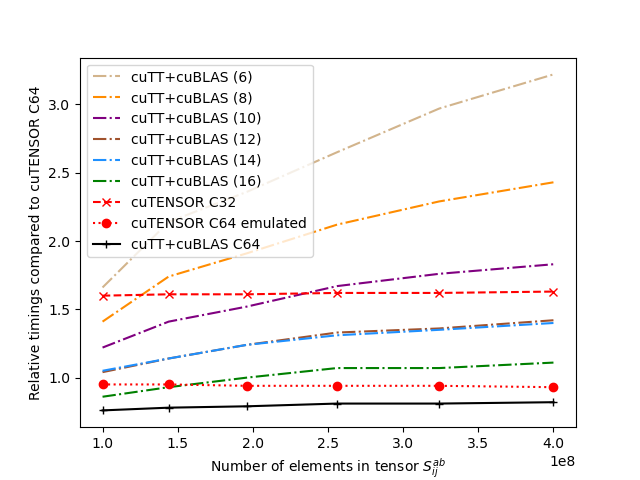}
  \end{minipage}
  \begin{minipage}[b]{0.75\textwidth}
    \includegraphics[width=\textwidth]{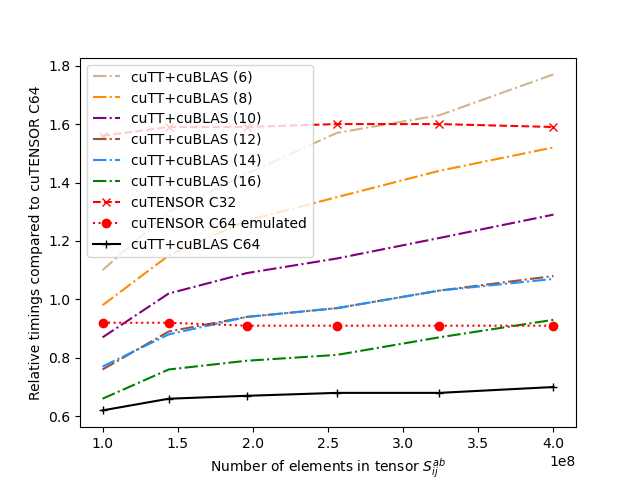}
  \end{minipage}
    \caption{OS-I TTGT (\texttt{cuBLAS}~\cite{cuBLAS} backend) relative timings with respect to the \texttt{cuTENSOR} C64 baseline using the native C64 compute datatype (values above 1 represent speedups) for PP (top) and PH (bottom) contractions, as a function of the number of elements in the result tensor $S_{ij}^{ab}$for precision \texttt{digits} varying from 6 to 16. We employed the \texttt{fix} and \texttt{eager} emulation parameters.\label{fig:speedups-vs-digits-cublas}}
\end{figure}

\begin{figure}[!tbp]
  \centering
  \begin{minipage}[b]{0.75\textwidth}
    \includegraphics[width=\textwidth]{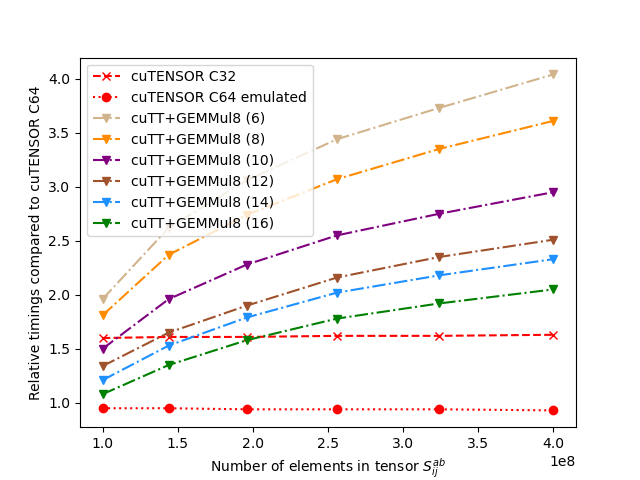}
  \end{minipage}
  \begin{minipage}[b]{0.75\textwidth}
    \includegraphics[width=\textwidth]{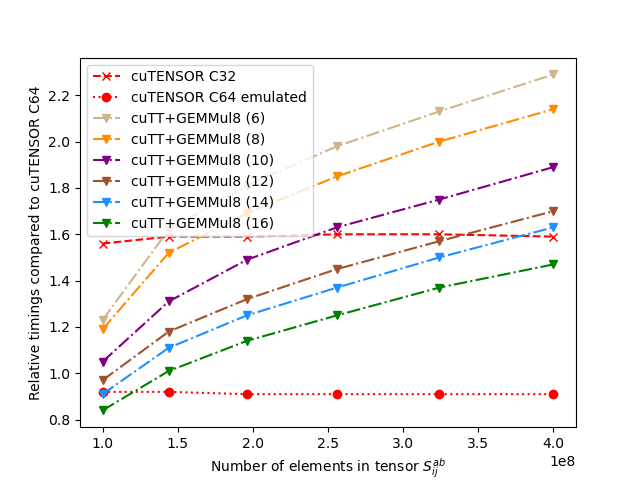}
  \end{minipage}
    \caption{OS-II TTGT (\texttt{GEMMul8} backend~\cite{OS-II-repo}) relative timings with respect to the \texttt{cuTENSOR} C64 baseline using the native C64 compute datatype (values above 1 represent speedups) for PP (top) and PH (bottom) contractions, as a function of the number of elements in the result tensor $S_{ij}^{ab}$\label{fig:speedups-vs-digits-gemmul8} for precision \texttt{digits} varying from 6 to 16. }
\end{figure}

In Figures~\ref{fig:speedups-vs-digits-cublas} and ~\ref{fig:speedups-vs-digits-gemmul8} we present additional comparisons of relative timings with respect to the \texttt{cuTENSOR} native C64 baseline~\cite{cuTENSOR-ref}, using the OS-I TTGT (\texttt{cuBLAS} backend) and the OS-II TTGT (\texttt{cuBLAS} backend) respectively.

From Figure~\ref{fig:speedups-vs-digits-cublas} we note, first, that already for 14 or 12 \texttt{digits}, the speedups for large contraction sizes approach the relative speedups of \texttt{cuTENSOR} C32 runs with respect to C64, even though the accuracy of emulated results is in principle higher. Second, we see that with 10, 8 or 6 \texttt{digits} it is possible to obtain substantial speedups for both PP and PH contractions compared to both \texttt{cuTENSOR}  C64 and C32 results. And third, we observe slight decreases in performance for \texttt{cuTENSOR}  C32 with increased contraction sizes for the PH contraction, whereas the OS-I TTGT scheme shows an improvement of performance with increased contraction size.

The results for the OS-II TTGT scheme, shown in Figure~\ref{fig:speedups-vs-digits-gemmul8} show qualitatively the same behavior as OS-I, with progressively improved performances as the number of \texttt{digits} is reduced. An important difference is that OS-II emulation consistently outperforms the reference, and consequently shows speedups compared to the baseline for significantly smaller problem sizes.

In all of these comparisons the \texttt{cuTENSOR} C32 reference uses native single precision compute; with BF16-based emulation~\cite{Bayraktar} the C32 curves would be expected to move further up, so they should not be read as an upper bound for single precision.

\clearpage
\section{Performance with respect to machine precision}
\begin{figure}[h]
    \centering
    \includegraphics[width=0.75\linewidth]{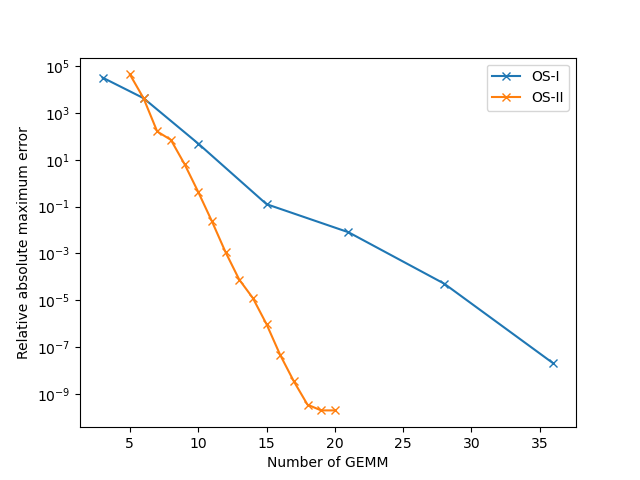}
    \caption{Error of OS-I and OS-II with respect to the number of GEMMs for input matrices A and B of shape 1024$\times$ 10000 and 10000$\times$1024 on A100. The OS-II has been tested for real arithmetic for this benchmark. The results are compared to the \texttt{cuBLAS} reference without any emulation.}
    \label{fig:OS-II-matrix-error}
\end{figure}

From the results presented in Figure 3 of the manuscript and the discussion from~\citet{Os-II-INT8}, we should expect the OS-II scheme to be better than the OS-I after a given number of GEMMs, since it has a linear convergence as opposed to the exponential convergence of the OS-I scheme on a logarithmic scale. We attempted to reproduce Figure 8 from~\citet{Os-II-INT8} with an A100 GPU card. For the configuration of this test, we generated the matrix elements using a random generator as in~\cite{Os-II-INT8} with $\phi=3$. From the results, presented in Figure~\ref{fig:OS-II-matrix-error}, we do not have the crossover at around 10 GEMMs but slightly before.

\clearpage
\bibliography{ms}